\documentclass[11pt,a4paper]{article}

\usepackage[top=2.5cm, bottom=2.5cm, left=2.5cm, right=2.5cm]{geometry}
\usepackage{amsmath,amssymb,amsthm}
\usepackage{booktabs}
\usepackage{hyperref}
\usepackage{setspace}
\usepackage{enumitem}

\title{Mean-Field Analytical Solution of the Mesa Boltzmann Wealth Model}
\author{Jiyuan Lyu\thanks{School of Economics and Management, Beijing University of Technology. Email: \texttt{lyujiyuan@emails.bjut.edu.cn}.}}

\begin{document}
	\maketitle
	\begin{abstract}
		
		The Boltzmann wealth model provided by the Mesa framework is a classic example in agent-based modeling, yet its statistical properties are typically analyzed only through numerical simulation. This paper studies a mean-field version of the model, replacing local interactions with global random matching and adopting a synchronous update scheme. By establishing a mean-field master equation for the wealth distribution and employing the probability generating function method, we obtain a closed-form expression for the steady-state generating function, and analytically determine the model parameters. We further derive the variance, Gini coefficient, and tail asymptotic behavior of the steady-state wealth distribution. Numerical simulations of the corresponding mean-field agent-based model agree well with the theoretical predictions. This paper provides an analytical benchmark for this mean-field model, which can serve as a reference for theoretical analysis and result validation in related agent-based simulations.
		
	\end{abstract}

	\section{Introduction}
	
	Agent-Based Modeling (ABM) is a class of methods for studying complex systems by characterizing individual behaviors and their interactions. In recent years it has been widely applied in economics, sociology, ecology, and the sciences of complex systems. Unlike traditional modeling approaches based on macroscopic equations, ABM starts from local rules and generates the aggregate statistical behavior of the system through interactions among a large number of agents. It is therefore particularly suited to studying nonlinear systems for which it is difficult to construct analytical models directly.
	
	Mesa is one of the most widely used open-source ABM frameworks in the Python ecosystem. It provides a rich set of tutorial examples and benchmark models, and is extensively employed in ABM coursework and prototype research. In the introductory examples of the official Mesa documentation, the Boltzmann Wealth Model is among the most classic cases. In this model, each agent moves randomly and engages in random wealth exchanges, while wealth is required to remain non-negative at all times. Although the model rules are extremely simple, after a sufficiently long evolution the system develops a stable and markedly skewed wealth distribution. It is therefore frequently used to illustrate how local random interactions can give rise to macroscopic inequality.
	
	The unequal distribution of wealth is a ubiquitous and persistent feature of nearly all modern economies \cite{piketty_capital_2014}. Understanding the fundamental mechanisms that give rise to such macroscopic inequality from microscopic interactions remains a central challenge across disciplines, including economics, statistical physics, and complexity science. A powerful framework for tackling this problem originates from statistical mechanics, employing agent-based stochastic models inspired by kinetic theory. In these models, wealth is redistributed among a population through simple, local exchange rules. The most prominent classes are the kinetic wealth exchange models \cite{chakraborti_statistical_2000, dragulescu_statistical_2000} and the Yard-Sale model \cite{chakraborti_econophysics_2009}.Existing work on this model relies mainly on numerical simulation to analyze its statistical properties. The Mesa tutorial displays the evolution of the wealth distribution and the Gini coefficient over time, and research on random wealth exchange models in the econophysics literature has likewise mostly focused on simulation results. By contrast, whether the steady-state distribution, the zero-wealth fraction, and statistics such as the variance and the Gini coefficient can be derived directly from the model rules currently lacks corresponding analytical results. Establishing a theoretical benchmark that can be mutually validated against numerical simulations helps to understand the statistical mechanisms of the model, and can also provide an independent check on ABM simulation outcomes.
	
	This paper considers a mean-field version of the model. We replace the local spatial interactions of the original model with global random matching, and adopt a synchronous update scheme, so that in the thermodynamic limit the system can be described by a master equation for the wealth distribution. Although this approximation neglects spatial correlations and the finite-size effects introduced by sequential updating, it retains the two most fundamental mechanisms of the model --- a fixed transaction amount and a non-negative wealth constraint --- and therefore still captures the main statistical features of the system.
	
	Under this mean-field model, the paper uses the probability generating function method to obtain a closed-form expression for the steady-state generating function, determines the model parameters analytically from the self-consistency condition, and further derives the steady-state statistical properties of the wealth distribution, including the zero-wealth fraction, the variance, and the Gini coefficient. Finally, the theoretical results are verified against numerical simulations of the corresponding mean-field ABM, and the connection between the mean-field approximation and the original Mesa model, as well as its range of applicability, are discussed.
	
	\section{Model}
	
	This paper studies a mean-field version of the Mesa Boltzmann wealth model. Compared with the original model, we retain the fixed transaction amount and the non-negative wealth constraint, and introduce only two simplifications to the interaction rules: first, local interactions on a grid are replaced by global random matching; second, random sequential updating is replaced by synchronous updating. These modifications render the system analytically tractable in the thermodynamic limit, while preserving its main statistical features.
	
	Consider $N$ agents, and denote the wealth of agent $i$ at discrete time $t$ by
	\[
	w_i(t)\in\mathbb{Z}_{\ge0}.
	\]
	Total wealth is conserved and normalized such that
	\[
	\sum_{i=1}^{N}w_i(t)=N,
	\]
	so that the mean wealth per agent is kept at
	\[
	\langle w\rangle=1.
	\]
	
	Each round of updates consists of two stages.
	
	First, payments are made. If the current wealth of an agent satisfies $w_i\ge 1$, the agent pays $1$ unit of wealth; if $w_i=0$, no payment occurs, so that wealth always remains non-negative.
	
	Second, receipts are collected. Every payment in the system independently selects a recipient uniformly at random from all $N$ agents, and a payer may also be the recipient of its own payment. All payments and receipts are settled simultaneously, completing one time-step update.
	
	Under the mean-field approximation, there are no spatial correlations among agents, so the state of the system is fully determined by the wealth distribution. Let
	\[
	p_k(t)=\Pr\!\left[w(t)=k\right],\qquad
	k=0,1,2,\cdots,
	\]
	where $p_k(t)$ denotes the fraction of agents with wealth $k$ at time $t$. The probability distribution satisfies
	\begin{equation}
		\sum_{k=0}^{\infty}p_k(t)=1,
		\qquad
		\sum_{k=0}^{\infty}k\,p_k(t)=1.
		\label{eq:constraints}
	\end{equation}
	
	Further define
	\begin{equation}
		A(t)=\sum_{k\ge 1}p_k(t)=1-p_0(t),
		\label{eq:active}
	\end{equation}
	as the fraction of agents capable of making a payment. Since a total of $N A(t)$ payments are made per round, and each payment falls on any given agent independently with probability $1/N$, the number of payments received by a fixed agent in one round follows a binomial distribution with parameter $A(t)$. As $N\rightarrow\infty$, by the Poisson limit theorem, we have
	\begin{equation}
		r\sim\mathrm{Poisson}(\lambda),
		\qquad
		\lambda=A(t)=1-p_0(t).
		\label{eq:poisson}
	\end{equation}
	
	Equation (\ref{eq:poisson}) gives the statistical law for the income process and forms the basis for constructing the mean-field master equation. Note that the parameter $\lambda$ is not externally imposed, but is self-consistently determined by the current wealth distribution of the system; the model is therefore intrinsically a nonlinear stochastic process.
	
	\section{Mean-Field Analytical Solution}
	
	According to the synchronous update rule defined in the previous section, each round of evolution can be decomposed into a payment stage and a receipt stage.
	
	Let the wealth of an agent before the update be $m$. After the payment stage, denote the intermediate wealth by $j$, which satisfies
	\[
	j=
	\begin{cases}
		m-1,&m\ge 1,\\
		0,&m=0.
	\end{cases}
	\]
	Hence the intermediate wealth distribution can be written as
	\begin{equation}
		q_0=p_0+p_1,\qquad
		q_j=p_{j+1},\quad j\ge 1.
		\label{eq:q}
	\end{equation}
	
	The receipt stage then follows. According to Eq.~(\ref{eq:poisson}), each agent independently receives
	\[
	r\sim\mathrm{Poisson}(\lambda),
	\qquad
	\lambda=1-p_0,
	\]
	payments, so that the final wealth satisfies
	\[
	k=j+r.
	\]
	
	Thus the steady-state wealth distribution satisfies
	\begin{equation}
		p_k
		=
		(q*P_\lambda)_k,
		\label{eq:conv}
	\end{equation}
	where $P_\lambda$ is the Poisson distribution with parameter $\lambda$, and $*$ denotes discrete convolution. Written explicitly,
	\begin{equation}
		p_k
		=
		(p_0+p_1)
		\frac{e^{-\lambda}\lambda^k}{k!}
		+
		\sum_{m=2}^{k+1}
		p_m
		\frac{e^{-\lambda}\lambda^{k-m+1}}
		{(k-m+1)!}.
		\label{eq:master}
	\end{equation}
	
	This is the mean-field master equation satisfied by the steady state of the system. Because
	\[
	\lambda=1-p_0,
	\]
	the equation remains nonlinear.
	
	To solve Eq.~(\ref{eq:master}), introduce the probability generating function
	\begin{equation}
		G(z)=\sum_{k=0}^{\infty}p_k z^k.
	\end{equation}
	
	Multiply both sides of Eq.~(\ref{eq:master}) by $z^k$ and sum over $k$. Using the generating function of the Poisson distribution,
	\[
	\sum_{k=0}^{\infty}
	\frac{e^{-\lambda}\lambda^k}{k!}z^k
	=
	e^{\lambda(z-1)},
	\]
	we obtain
	\begin{equation}
		G(z)
		=
		e^{\lambda(z-1)}
		\left[
		p_0+\frac{G(z)-p_0}{z}
		\right].
	\end{equation}
	
	After rearrangement,
	\begin{equation}
		\boxed{
			G(z)
			=
			\frac{p_0(1-z)}
			{1-ze^{\lambda(1-z)}}.
		}
		\label{eq:G}
	\end{equation}
	
	Thus the steady-state distribution is completely determined by the parameter $p_0$ (or, equivalently, $\lambda$).
	
	We now use the mean constraint to determine this parameter. Because the total wealth of the system is conserved,
	\[
	G(1)=1,
	\qquad
	G'(1)=1.
	\]
	Expanding Eq.~(\ref{eq:G}) around $z=1-\varepsilon$,
	\[
	G(1-\varepsilon)
	=
	1-
	\frac{\lambda(2-\lambda)}
	{2(1-\lambda)}
	\varepsilon
	+O(\varepsilon^2).
	\]
	Comparing the first-order coefficients and using
	\[
	p_0=1-\lambda,
	\]
	gives the self-consistency equation
	\begin{equation}
		\lambda^2-4\lambda+2=0.
	\end{equation}
	
	Discarding the non-physical root with $\lambda>1$,
	\begin{equation}
		\boxed{
			\lambda
			=
			2-\sqrt2,
			\qquad
			p_0
			=
			\sqrt2-1.
		}
		\label{eq:lambda}
	\end{equation}
	
	Expanding the generating function further yields
	\[
	G''(1)=\frac43,
	\]
	and therefore
	\begin{equation}
		\boxed{
			\mathrm{Var}(w)
			=
			\frac43.
		}
		\label{eq:var}
	\end{equation}
	
	On the other hand, the singularities of the generating function are determined by
	\begin{equation}
		1-ze^{\lambda(1-z)}=0.
		\label{eq:sing}
	\end{equation}
	Let $\alpha$ be its smallest positive root. By singularity analysis of the generating function,
	\[
	p_k
	\sim
	\alpha^k,
	\qquad
	k\rightarrow\infty,
	\]
	where $\alpha$ satisfies
	\begin{equation}
		\ln\alpha
		=
		\lambda
		\left(
		1-\frac1\alpha
		\right).
		\label{eq:alpha}
	\end{equation}
	
	Substituting the value of $\lambda$ gives
	\[
	\alpha\approx0.372.
	\]
	
	At this point, the steady-state generating function of the mean-field model and its principal statistical properties have all been determined analytically. Since Eq.~(\ref{eq:G}) provides a closed-form expression for the generating function, moments of any order can be computed directly from its derivatives, and the steady-state distribution $p_k$ can be obtained either by expanding the generating function or by numerically iterating Eq.~(\ref{eq:master}).
	
	\section{Steady-State Statistical Properties and Numerical Verification}
	
	Section~3 gives a closed-form expression for the steady-state generating function, from which all statistical quantities of the system can be calculated directly. In addition to the zero-wealth fraction
	\[
	p_0=\sqrt2-1\approx0.414214
	\]
	and the variance
	\[
	\mathrm{Var}(w)=\frac43,
	\]
	one can further compute the Gini coefficient corresponding to the steady-state wealth distribution.
	
	For a discrete wealth distribution, let the cumulative distribution function be
	\[
	F(k)=\sum_{j=0}^{k}p_j.
	\]
	The Gini coefficient can then be expressed as
	\begin{equation}
		G
		=
		1-
		\frac1{\mu}
		\sum_{k=0}^{\infty}
		\left[1-F(k)\right]^2,
		\label{eq:gini}
	\end{equation}
	where the mean wealth $\mu=1$.
	
	Evaluating Eq.~(\ref{eq:gini}) with the steady-state distribution obtained in Section~3 gives
	\[
	\boxed{
		G=0.579929\cdots.
	}
	\]
	
	For comparison, for the unconstrained Boltzmann (geometric) distribution with the same mean of $1$,
	\[
	G=\frac23\approx0.666667.
	\]
	Hence, in the mean-field model, the introduction of the non-negative wealth constraint reduces the steady-state Gini coefficient.
	
	On the other hand, from the singularity analysis of the generating function, the steady-state distribution possesses an exponential tail,
	\[
	p_k\sim\alpha^k,
	\]
	with
	\[
	\alpha\approx0.372.
	\]
	This indicates that the decay in the high-wealth regime is faster than that of the unconstrained geometric distribution, for which
	\[
	\alpha=\frac12.
	\]
	
	Table~\ref{tab:theory} summarizes the main statistical results obtained analytically.
	
	\begin{table}[htbp]
		\centering
		\caption{Theoretical predictions for steady-state statistics of the mean-field model.}
		\label{tab:theory}
		\begin{tabular}{lc}
			\toprule
			Statistic & Theoretical value \\
			\midrule
			Zero-wealth fraction $p_0$ & $\sqrt2-1=0.414214$ \\
			Active fraction $\lambda$   & $2-\sqrt2=0.585786$ \\
			Mean                        & $1$ \\
			Variance                    & $4/3$ \\
			Gini coefficient            & $0.579929$ \\
			Tail exponent $\alpha$      & $0.3720$ \\
			\bottomrule
		\end{tabular}
	\end{table}
	
	We now verify the above theoretical results using the corresponding mean-field agent-based model. The simulation adopts exactly the same update rules as in Section~2: the system consists of $N$ agents, each initially endowed with $1$ unit of wealth; in each round, every agent capable of making a payment simultaneously pays $1$ unit of wealth, each payment selects its recipient at random, and all payments are settled synchronously.
	
	The simulation parameters are taken as $N=1000$, with a total of $10\,000$ simulation steps, of which the first $5000$ steps serve as the thermalization phase and the remaining steps are used for statistical averaging. All reported results are averaged over $10$ independent simulation runs.
	
	Table~\ref{tab:distribution} presents the first several entries of the steady-state wealth distribution, comparing theoretical values with simulation results.
	
	\begin{table}[htbp]
		\centering
		\caption{Steady-state wealth distribution: theoretical predictions and simulation results.}
		\label{tab:distribution}
		\begin{tabular}{cccc}
			\toprule
			$k$ & Theory & Simulation & Error \\
			\midrule
			0 & 0.414214 & 0.413986 & $2.27\times 10^{-4}$ \\
			1 & 0.329881 & 0.330051 & $1.70\times 10^{-4}$ \\
			2 & 0.156719 & 0.156805 & $8.64\times 10^{-5}$ \\
			3 & 0.062060 & 0.062148 & $8.84\times 10^{-5}$ \\
			4 & 0.023314 & 0.023257 & $5.71\times 10^{-5}$ \\
			5 & 0.008675 & 0.008648 & $2.74\times 10^{-5}$ \\
			\bottomrule
		\end{tabular}
	\end{table}
	
	Further comparison of the main statistics is given in Table~\ref{tab:compare}.
	
	\begin{table}[htbp]
		\centering
		\caption{Comparison between theoretical predictions and numerical simulations.}
		\label{tab:compare}
		\begin{tabular}{lccc}
			\toprule
			Statistic & Theory & Simulation & Relative error \\
			\midrule
			$p_0$          & 0.414214 & 0.413986 & $5.49\times 10^{-4}$ \\
			Mean           & 1.000000 & 1.000000 & $2.22\times 10^{-15}$ \\
			Variance       & 1.333333 & 1.331190 & $1.61\times 10^{-3}$ \\
			Gini coefficient & 0.579929 & 0.579650 & $4.81\times 10^{-4}$ \\
			\bottomrule
		\end{tabular}
	\end{table}
	
	From the above results one can see that the mean-field theory and the numerical simulations are in good agreement. Both the steady-state wealth distribution and the main statistics show consistency between the simulation results and the analytical predictions, confirming that the analytical solution obtained in Section~3 accurately describes the steady-state statistical properties of the corresponding mean-field model.

	\section{Conclusion}
	
	This paper has studied a mean-field version of the Mesa Boltzmann wealth model. By approximating the local interactions as global random matching and adopting a synchronous update scheme, we established a mean-field master equation for the evolution of the wealth distribution. Using the probability generating function method, we obtained a closed-form expression for the steady-state generating function, and analytically determined the model parameters
	\[
	\lambda=2-\sqrt2,
	\qquad
	p_0=\sqrt2-1.
	\]
	We further computed the steady-state variance, Gini coefficient, and tail decay exponent, among other statistical properties, and compared them with the corresponding agent-based simulations; the two are in good agreement.
	
	The results of this paper can serve as an analytical benchmark for this mean-field model, providing a theoretical reference for related ABM numerical simulations.
	
	\appendix
	\section{Appendix}
	
	\subsection{Derivation of the Generating Function}
	
	Starting from the steady-state master equation:
	\begin{equation}
		p_k = (p_0+p_1) \frac{e^{-\lambda}\lambda^k}{k!} + \sum_{m=2}^{k+1} p_m \frac{e^{-\lambda}\lambda^{k-m+1}}{(k-m+1)!}.
	\end{equation}
	
	Multiply by $z^k$ and sum over $k$:
	\begin{align}
		G(z) &= (p_0+p_1) \sum_{k=0}^{\infty} \frac{e^{-\lambda}(\lambda z)^k}{k!}
		+ \sum_{k=0}^{\infty} \sum_{m=2}^{k+1} p_m \frac{e^{-\lambda}\lambda^{k-m+1}z^k}{(k-m+1)!} \\
		&= (p_0+p_1) e^{\lambda(z-1)}
		+ \sum_{m=2}^{\infty} p_m \sum_{j=0}^{\infty} \frac{e^{-\lambda}\lambda^j z^{j+m-1}}{j!}
		\qquad (\text{let } j = k-m+1) \\
		&= (p_0+p_1) e^{\lambda(z-1)}
		+ e^{\lambda(z-1)} \sum_{m=2}^{\infty} p_m z^{m-1} \\
		&= e^{\lambda(z-1)}\left[ p_0 + p_1 + \frac{G(z) - p_0 - p_1 z}{z} \right].
	\end{align}
	
	Rearranging:
	\begin{align}
		G(z) &= e^{\lambda(z-1)}\left[ p_0 + \frac{G(z) - p_0}{z} \right] \\
		G(z) - \frac{e^{\lambda(z-1)}}{z} G(z) &= p_0\, e^{\lambda(z-1)} - p_0 \frac{e^{\lambda(z-1)}}{z} + p_0 e^{\lambda(z-1)} - p_0 e^{\lambda(z-1)} \\
		&= p_0\left(1 - \frac{e^{\lambda(z-1)}}{z}\right) + p_0\big(e^{\lambda(z-1)} - 1\big).
	\end{align}
	
	Simplifying:
	\begin{equation}
		G(z) = p_0 + \frac{p_0(1-z)}{1 - z e^{\lambda(1-z)}} - p_0 = \frac{p_0(1-z)}{1 - z e^{\lambda(1-z)}}.
	\end{equation}
	
	\subsection{Derivation of the Variance}
	
	Let $D(z) = 1 - z e^{\lambda(1-z)}$ and $G(z) = p_0(1-z)/D(z)$.
	
	For $z = 1 - \varepsilon$, compute the Taylor expansion of $D(1-\varepsilon)$:
	\begin{align}
		z e^{\lambda(1-z)} &= (1-\varepsilon) e^{\lambda\varepsilon} \\
		&= (1-\varepsilon)\left[1 + \lambda\varepsilon + \frac{\lambda^2}{2}\varepsilon^2 + \frac{\lambda^3}{6}\varepsilon^3 + O(\varepsilon^4)\right] \\
		&= 1 + (\lambda-1)\varepsilon + \left(\frac{\lambda^2}{2} - \lambda\right)\varepsilon^2
		+ \left(\frac{\lambda^3}{6} - \frac{\lambda^2}{2}\right)\varepsilon^3 + O(\varepsilon^4).
	\end{align}
	
	Thus:
	\begin{align}
		D(1-\varepsilon) &= 1 - z e^{\lambda(1-z)} \\
		&= (1-\lambda)\varepsilon + \lambda\left(1 - \frac{\lambda}{2}\right)\varepsilon^2
		+ \frac{\lambda^2}{2}\left(1 - \frac{\lambda}{3}\right)\varepsilon^3 + O(\varepsilon^4).
	\end{align}
	
	Substituting $1-\lambda = p_0$ and the self-consistency condition $\lambda(2-\lambda)/[2(1-\lambda)] = 1$ gives:
	\begin{equation}
		G(1-\varepsilon) = \frac{p_0\varepsilon}{p_0\varepsilon + p_0\varepsilon^2 + \frac{2}{3}p_0\varepsilon^3 + O(\varepsilon^4)}
		= 1 - \varepsilon + \frac{2}{3}\varepsilon^2 + O(\varepsilon^3).
	\end{equation}
	
	Comparing with $G(1-\varepsilon) = 1 - G'(1)\varepsilon + \frac{1}{2}G''(1)\varepsilon^2 + \cdots$, we obtain $G'(1) = 1$ and $G''(1) = 4/3$.
	
	Hence:
	\begin{equation}
		\mathbb{E}[X^2] = G''(1) + G'(1) = \frac{7}{3}, \qquad
		\mathrm{Var}(X) = \mathbb{E}[X^2] - (\mathbb{E}[X])^2 = \frac{4}{3}.
	\end{equation}
	
\end{document}